\documentclass[epj]{svjour}
\usepackage{latexsym}
\usepackage[dvips]{graphicx}

\begin{document}
\title{Very low shot noise in carbon nanotubes}
\author{P.-E.~Roche\inst{1,2}, M. Kociak\inst{1}, S.
Gu\'eron\inst{1}, A.Kasumov\inst{1,3}, B. Reulet\inst{1} \and H.
Bouchiat\inst{1} }
\institute{Laboratoire de Physique des Solides, Associ\'e au CNRS,
B\^atiment 510, Universit\'e Paris-Sud, 91405 Orsay, France
\and Laboratoire de Physique de la Mati\`ere Condens\'ee, Ecole
Normale Sup\'erieure, 24 rue Lhomond, 75231 Paris Cedex 05, France
\and Institute of Microelectronics Technology and High Purity
Materials,Russian Academy of Sciences, Chernogolovka 142432 Moscow
Region, Russia
}
\date{ }
%
\abstract{We have performed noise measurements on suspended ropes of
single wall carbon nanotubes (SWNT) between 1 and 300~K for different
values
of dc current through the ropes. We find that the shot noise is
suppressed by more than a factor 100 compared to the full
shot noise 2eI. We have also measured an
individual SWNT and found a level
of noise which is smaller than the minimum expected. Another finding
is the
very low level of 1/f noise, which is significantly lower than
previous observations.
We propose two possible interpretations for this strong shot noise
reduction: i) Transport within a rope takes place through a  few nearly
ballistic tubes within a rope  and possibly involves non integer effective
charges  with $ e^*\sim 0.3 e$. ii) A substantial fraction of the tubes
conduct with a strong reduction of effective charge
(by more than a factor~50).
} 
\maketitle
\section{Introduction}

Metallic single wall carbon nanotubes (SWNT) are known to be model
systems for the study of one dimensional(1D) electronic transport. Depending on
diameter and helicity, SWNT have at most 2 conducting channels and in
the absence of disorder their minimum resistance is predicted to be
$h/4e^2\approx~6.5~k\Omega$ \cite{Dresselhaus,Hamada}. In 1D,
electronic interactions are expected to lead to a breakdown of the
Fermi
   liquid state. Nanotubes should then be described by Luttinger
Liquids (LL) theories~\cite{Kane,egger}, with collective low energy
   excitations.
Indications of the validity of LL description with repulsive
interactions in SWNT were given by the measurement of a resistance
diverging
as a power law with temperature down to 10~K~\cite{bockrath}. It has
been shown that another experimental signature of a Luttinger liquid
is the existence of non integer charge excitations $e^* =ge$ where g
is the Luttinger parameter which depends on the interaction strength
( g=1 in the absence of interactions.) These excitations can in
principle be detected in a 2-wire measurement on a sample containing
an impurity in the weak
backscattering regime~\cite{blanter,kanefrac}. Indeed, non integer
charge excitations create a current noise which has the form:

\begin{equation}
S_I =2e^*i_b
	\label{eqI}
	\end{equation}
where $i_b$ is the back-scattered current. Edge states in 2 D electron
systems in the fractional quantum Hall regime have provided an
elegant demonstration of this prediction~\cite{exp}.
However carbon nanotubes, contrary to the edge states in the
fractional quantum hall regime, are expected to constitute a non
chiral Luttinger Liquid
where forward and backward electrons are not spatially separated.
Theoretical predictions for the non chiral LL are quite controversial.
Although the initial prediction eq.\ref{eqI}~\cite{kanefrac} was
obtained for a non chiral infinite LL, it has been recently
claimed~\cite{jap,trauz}
that the shot noise in a LL between two reservoirs does not reveal
the quasi-particle non integer charge and the authors in~\cite{trauz}
predict
instead an integer charge of 1e, in the expression of shot noise.
These calculations however rely on drastic hypothesis concerning
screening of electron interactions and boundary conditions with the
reservoirs which validity can be questioned~\cite{lederer}.
The investigation of shot noise in carbon nanotubes is therefore an
important issue.
The measurements we present on SWNT show a surprisingly high shot
noise reduction.

\section{Experimental set-up}

The SWNT are prepared by an electrical arc method with a mixture of
nickel and yttrium as a catalyst~\cite{Journet,Vaccarini}.
SWNT with diameters of the order of 1.4 nm are obtained. They are
purified by the cross-flow filtration method~\cite{Vaccarini}.
The tubes usually come assembled in ropes of a few hundred parallel
tubes, but individual tubes can also be obtained after chemical
   treatment with a surfactant~\cite{burghard}. Isolation of an
individual rope and connection to measuring pads are performed
according
to the following nano-soldering technique.
A target covered with nanotubes is placed above a
metal-coated~\cite{MetalCoating} suspended $Si_{3}N_{4}$ membrane,
in which a roughly  1~ $\mu m$ by 100~ $\mu m$ slit has been etched 
with a focused ion beam.
Following ref.~\cite{Kasumov2} a focused UV laser beam pulse (power
10 kW) is applied for 10 ns,
a nanotube drops from the target and connects the edges of the slit.
Since the metal electrodes on each side of the slit are locally
molten,
the tube gets soldered into good contact.
The resulting samples can then be characterized using transmission
electronic microscopy. The measurement of large supercurrents when
the tubes
are soldered to superconducting electrodes \cite{Kasumov} indicate 
that the contact
resistance is much smaller than the intrinsic resistance of the tubes.

Using this technique, we have obtained individual carbon nanotubes such as
the sample presented in fig.(5),
with resistances as low as $10~k\Omega$, that present a quasi metallic
behavior down to very low temperature with less than a $25\%$ increase of
resistance between 300~K and 1~K. Ropes containing few tens of nearly
ordered carbon nanotubes soldered using the same technique present a
wide range of measured
resistances that can vary between $100$ and $10^5 ~\Omega$ at 300~K.
These ropes are generally metallic when their resistances are below
$10~k\Omega$ at room temperature.
The data presented here are taken on ropes that have a resistance
less than 1~ $k\Omega$. (More resistive ropes usually had a level of
1/f noise that prevents
the measurement of shot noise.) The temperature
dependence of these
low resistance ropes (see inset of fig. 1) is also very weak but in
contrast with what is observed in individual tubes, it is not
monotonous as already
reported in~\cite{fisher}. The resistance decreases linearly as
temperature is lowered between room temperature
and 30~K indicating the freezing-out of  phonon modes, and then
increases as T is further decreased, just as in
individual tubes.

The noise was recorded as a voltage fluctuation $\Delta V(t)$ across
the samples  biased with various DC currents $I_{DC}$.
The spectrum of voltage noise power

$P_V(f) =2\int_{}^{}\overline{\Delta V(t)\Delta V(O)}\exp(2i\pi f t)
dt$ was averaged during over roughly 10 minutes and
converted into a current noise according to $P_{I}=h.P_{V}/R^2$,
where $R$ is the sample resistance and $h= (1+R/Z)^2$
is a correction factor which accounts for the finite impedance $Z>>R$
of the biasing circuit. For all the samples tested, this correction
factor
never exceeds the value $1.2$. In practice, a spectrum analyser
calculated the noise power spectrum by correlating the output signals
from two low noise differential preamplifiers (model LI-75A by NF
Electronic Instruments) which were independently connected across the
samples.
The measuring procedure has been checked by
replacing the samples with a $560~\Omega$ macroscopic resistor, for
which zero shot noise is expected.
The experiment, conducted at 4.2~K, gave the expected result within
$\pm 2eI/2000$ (see fig. 2).
Another validation of the apparatus and correlation procedure was
provided by the measurement of the full shot noise 2eI across the PN
junction
of a commercial transistor at room temperature.

\begin{figure}
    \begin{center}
      \includegraphics[width=8cm]{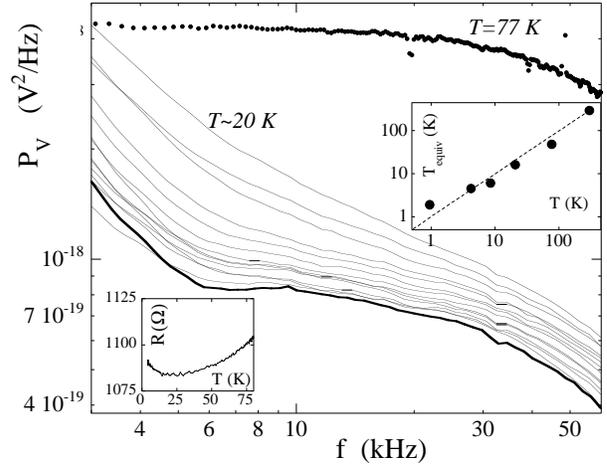}
    \end{center}
    \caption{Typical voltage noise power spectra $P_{V}$ at 77~K (zero
current bias ) and 20.6~K ($I_{DC}=0$ (thick line)$ , 20, 40, \ldots,
260~\mu A$) for the Rope A.
    The high frequency cut-off is due to a protective low pass
filtering. The dashed lines are the expected Johnson Nyquist noise
floor.
    Lower Inset:Temperature dependence of the resistance of Rope A.
    Upper Inset: Noise floor for $I_{DC}=0$ in~K ($P_{V}/4kR$) versus
temperature.
    }
    \label{fig1}
\end{figure}

\section{Results}

Typical noise spectra obtained on a rope of 200 tubes, which
resistance varies between $1085~\Omega$ and $1100~\Omega$ are shown
on fig.\ref{fig1} for $I_{DC}$ currents up
to $260~\mu A$ (Rope A). On all spectra, we have eliminated
parasitic peaks (due to harmonics of 50 Hz and radiations of monitors
and screens),
which are independent of $I_{DC}$. They represent about $1\%$ of the
investigated frequencies.
To increase the readability of the fig.1, the data used for the plot
have been smoothed over a 600 Hz frequency window.
One can note also the high frequency cutoff above 30 kHz which is due
to input filters in the experiment.

\begin{figure}
    \begin{center}
      \includegraphics[width=8cm]{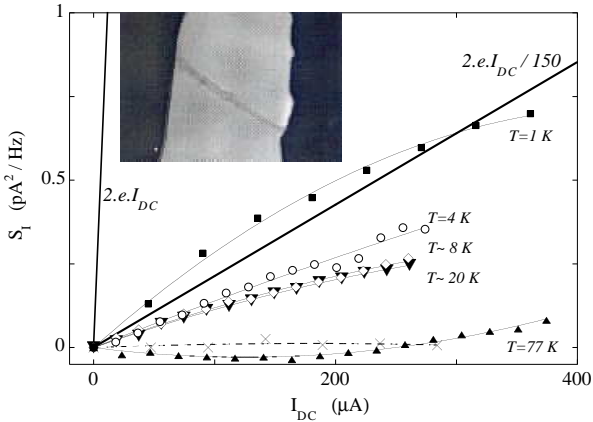}
    \end{center}
    \caption{
    Shot noise $S_{I}=P_{I}(I_{DC})-P_{I}(O)$ versus bias current
$I_{DC}$ for the Rope A at different temperatures. The thick solid
lines represent the
    1/150 reduced and full shot noise $2eI_{DC}$. X : shot noise of a 
$560~\Omega$
macroscopic resistor at 4.2~K (control sample).
   Inset : TEM micrograph of the Rope A, from which we deduce its
length $L=0.4~\mu m$ and the number of tubes $N$ from the measured
diameter $D$
    through $N= (D/(d+e))^2$, where d is the diameter of a single tube
(d=1.4 nm), and e is the typical distance between tubes in a rope
(e=0.2 nm).
    }
    \label{fig2}
\end{figure}

Every sample exhibits three intrinsic noise contributions : Johnson
Nyquist thermal noise, shot noise and low frequency noise. In
addition,
the major sources of background noise are the circuit resistances and
preamplifiers voltage and current noise.
These different parasitic contributions have been
characterized independently with specific experiments, in particular
at 140~mK in order to reduce the thermal noises.
For all measurements, the total background noise is dominated by the
amplifiers current noise ($130~fA.Hz^{-0.5}$ at 10 kHz) and it always
remains
smaller than the noise from the sample itself.

For $I_{DC}=0$ and beyond the low frequency region, the white noise
$P_{I}^{0}$ is identified with Johnson Nyquist noise. Its level was
determined after correction for the high frequency cutoff, which is
due to knwon filter capacitances. This noise
power can be expressed as an
equivalent temperature $T_{eq}$ according to
$T_{eq}=P_{I}^{0}R/4k_B$. The upper inset of fig.1 shows that $T_{eq}$ is
always close to the measured value of T,
which provides a basic experimental check.

The quantity we consider from now on is the difference between the
noise power spectra obtained with and without DC current bias :
$P_{I}(I_{DC})-P_{I}(O)$.
This difference only contains shot noise and low frequency noise.
We find that the low frequency noise roughly scales as $I^2/f$, which
is the basic signature of 1/f noise originating from the a population
of active two-level systems in a metallic conductor~\cite{dutta}.

Note that the amplitude of 1/f noise we measure is very small
compared to the values already published on carbon
nanotubes~\cite{paalanen,postma}.
A possible explanation is the suspended character of our samples in
contrast with deposited nanotubes which can interact with residual
mobile
charges present on any substrate. Following the analysis
in~\cite{postma}we characterize the amplitude of 1/f noise by the
dimensionless
coefficient B defined as $S_I/I_{dc}^2 =B/f$.  For rope A,
we find $B~\simeq 10^{-13}$ nearly
independent of temperature between 300~K and 1~K.
   For the highest currents (typically $I_{DC}>350~\mu A$), an
additional low frequency noise appears with a $1/f^2$ dependence
which saturates at very low frequency. This behavior is typical of
telegraphic noise which characteristic frequency
increases with $I_{DC}$ (see fig. 1) and is similar to previous
observations in nanometer scales metallic samples where the
resistance noise is dominated
by the activity of an individual or a very few number of
two-level-systems.\cite{ralls}. We fit the tail of the low frequency
noise by a $1/f^\alpha$ power law, and the remaining
white noise is identified to the shot noise $S_{I}$. Above 5-10 $kHz$,
the amplitude of the white noise level is not found to depend on the
details of the low frequency fit, at least for $\alpha$ arbitrarily
chosen between 1 and 2.

We now focus exclusively on this shot noise. In fig. 2, $S_{I}$ is plotted
as a function of $I_{DC}$ for Rope A. As expected we observe
a roughly linear dependence. Interestingly,
the current noise is strongly reduced (by more than a factor 150)
compared to the full shot noise value $S_I =2eI$. For this reason it
cannot be detected
for currents below $10~\mu A$. This strong reduction of
shot noise is the central result of our paper. Note however that we
cannot discriminate
in these experiments the intrinsic noise of the tubes from the
noise of the contacts. It is thus possible that the intrinsic noise
of the nanotubes
is even smaller than these measured values.

This small shot noise amplitude decreases systematically as
temperature is increased between 1 and 20~K and is not measurable 
anymore at 77~K
just like in a
macroscopic resistor. Shot noise is indeed expected to disappear at
temperatures for which inelastic electron phonon scattering
takes place inside the sample, with scattering lengths much smaller
than the sample length. (And indeed the increase of resistance
measured above 30~K indicates
that e-phonon scattering takes place inside the sample at this temperature).

Similar shot noise reductions where also seen (see fig.3) in the less resistive
ropes B ($R=495~\Omega$) and C ($R=186~\Omega$)  where reduction 
factors are respectively measured to be 270 and 150 at 1K.
We have also performed noise measurements on an individual tube at
1~K (fig. 4)which resistance varies between 10.5 and $12.5~ k\Omega$
between $300~K$ and $1~K$. The amplitude of 1/f noise measured in 
this sample is
much higher than in
ropes, and presents strong non-linearities with $I_{dc}^2$ above 
$0.1\mu A$as shown on
fig.5. At current below $0.1~\mu A$ it can be described by the
coefficient $B=10^{-9}$ which is about
$10^2$ smaller than the value recently measured at 8~K in an
individual tube 5 times more resistive by Postma et {\it
al.}~\cite{postma}.
The strong difference by more than a factor $10^4$ of the amplitude
of 1/f noise between ropes containing 300 tubes and an individual
tube may seem surprising.
But it can be qualitatively understood considering that transport
through a rope essentially probes the most conducting tubes which are
also likely to be the less noisy ones.
This large amount of 1/f noise in the individual tube sample
makes the analysis of the shot noise component more difficult. The
data shown in fig. 5 shows that there is barely no detectable shot noise in
this sample.
We can nevertheless give an upper-bound estimate of the amount of
  shot noise which are the data points shown.

\begin{figure}
    \begin{center}
      \includegraphics[width=8cm]{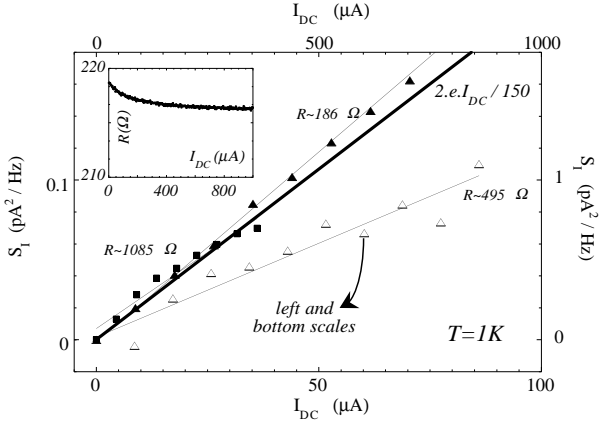}
    \end{center}
    \caption{
    Shot noise $S_{I}$ versus bias current $I_{DC}$ at 1~K for the
different Ropes A (full squares), B (empty triangles) and C (full
triangles) which have
    similar lengths and diameters. Note different scales both on Y and X
axis for the rope B with $R=495~\Omega$. The thick solid line
represents
    the reduced shot noise $2eI_{DC}/150$ on both scales. Inset:
Differential resistance versus applied dc current for the rope C. }

    \label{fig3}
\end{figure}

\begin{figure}
    \begin{center}
      \includegraphics[width=8cm]{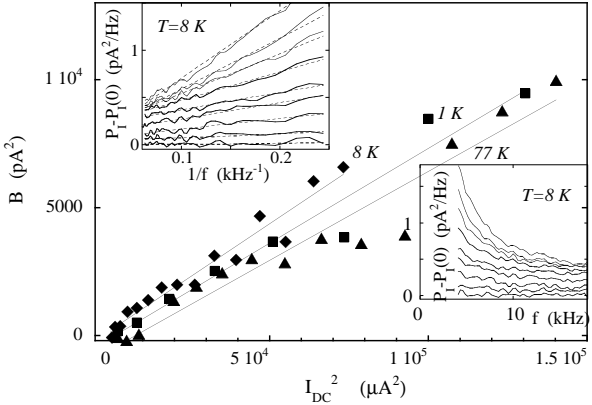}
    \end{center}
    \caption{
    $1/f$ noise in Rope A. Proportionality constant between the low 
frequency excess
noise and (1/f) versus $I_{DC}^2$, at different temperatures.
    Lower inset : Low frequency excess noise $P_{I}(I_{DC})-P_{I}(O)$
versus frequency at $T=8~K$.
    Upper inset : Low frequency excess noise versus 1/frequency at
$T=8~K$. The solid lines are linear fits for $f<16~kHz$ and
correspond to $B=10^{-13}$.
    }
    \label{fig4}
\end{figure}

\begin{figure}
     \begin{center}
       \includegraphics[width=8cm]{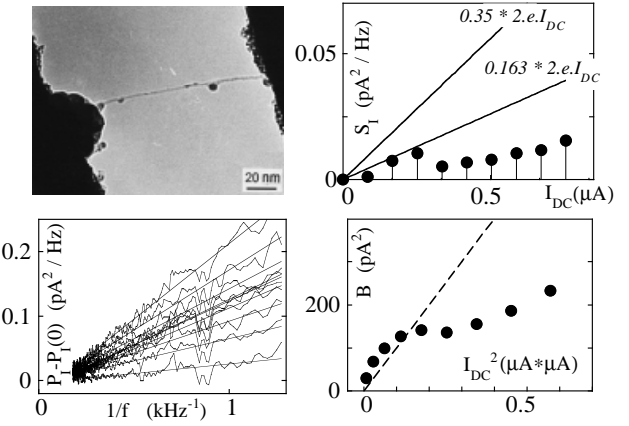}
     \end{center}
     \caption{
     Individual tube at T=1~K.\textit{ Upper-left inset:} TEM
micrograph of the sample which is $0.15~\mu m$ long.
     \textit{Upper-right inset:} Shot noise
$S_{I}=P_{I}(I_{DC})-P_{I}(O)$ versus bias current $I_{DC}$ at 1~K
for the individual tube. The solid circles
     correspond to the maximal shot noise compatible with the data, as
indicated by the error bars.
     Note however that the residual background noise of
$P_{I}(O)=0.04~pA^{2}/Hz$ measured for this sample is 2.5 times higher
than the expected thermodynamic noise at 1~K.
     The solid lines are the lower and upper bounds for the value of
$S_{I}$ for a two independent channel conductor compatible with the
sample resistance $R=10.5 k\Omega$ measured at room temperature.
     They correspond to reduction factor of 0.163 and 0.350.
    \textit{ Lower-left inset:} Low frequency excess noise
$P_{I}(I_{DC})-P_{I}(O)$ versus 1/frequency. The solid lines are
linear fits for $f<5.5 kHz$.
    \textit{Lower-right inset:} Amplitude of low frequency excess (1/f)
noise versus $I_{DC}^2$. The solid line is a linear fit only valid in
the low
    current region from which the coefficient $B=10^{-9}$ is extracted).
     }
     \label{fig5}
\end{figure}

\section{Discussion}

We now discuss the possible physical mechanisms which could be at the
origin of the strong shot noise reduction.

i) Strong electron phonon scattering in the sample or at the contacts.
Electron phonon scattering characterized by a typical scattering
length $l_{eph}$ has been shown to induce a shot noise reduction of a
factor
$l_{eph}/L$~ for a sample of length L much longer than
$l_{eph}$ \cite{wees88}. However in order to explain the noise reduction factor
observed in
the present experiments an inelastic length as short as 5 $nm$ is
needed which is quite unphysical at 1~K and at low current.
Observation of proximity
induced superconductivity in the same type of samples is a
confirmation of long inelastic scattering lengths in the tubes and
absence
of inelastic scattering at the contacts. Of course,
this electron phonon scattering is expected to increase at high currents,
as has been invoked
to explain the strong non-linearities observed in low temperature transport in
individual tubes observed above $I=5~\mu A$ per tube~\cite{yao}.
We have checked that non linearities
for all measured samples discussed here remain extremely small: less
than 1 percent
of resistance (see inset of fig. 3) and are opposite in sign compared
to what is expected for
current-induced electron phonon scattering where R should exhibit an
increase with current instead of the small decrease observed here.

ii) Ballistic transport through the tubes in a Landauer type of
picture~\cite{landauer}.
Assuming that carbon nanotubes in a rope behave as independent 2
channel conductors characterized by their transmissions $t_i^a$ and
$t_i^b$,
the total conductance of a rope reads :

\begin{equation}
G = G_0 \Sigma_i ( t_i^a+ t_i^b)
\end{equation}
where $G_0 =2 e^2/h$ is the conductance quantum. The shot noise
spectral density through the sample biased with the voltage V in the
limit $eV >> k_BT$ is:
\begin{equation}
  S_I = 2G_0 e V \Sigma_i \left [t_i^a(1- t_i^a)+t_i^b(1- t_i^b) \right ]
\end{equation}

\begin{figure}
     \begin{center}
       \includegraphics[width=8cm]{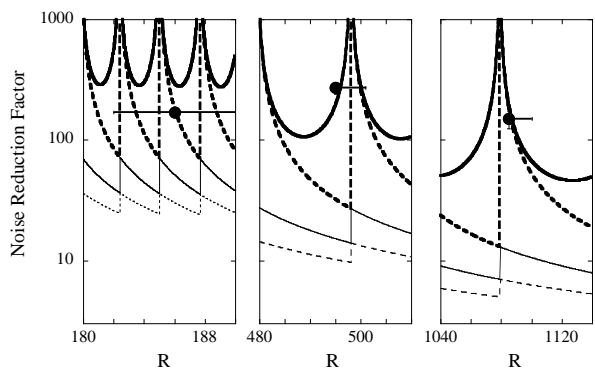}
     \end{center}
     \caption{
     Noise reduction factor for a multichannel conductor of
resistance R (axis) for 2 different types of transmission
distribution.
Thick solid line: all conducting channels
have a full transmission except one. This corresponds to the maximum reduction
attainable considering all the combinations of transmissions.
Other lines: all conducting channels have the same transmission 
$t_{i}=\mathcal{G}/N_{G}$.
Thick dashed line: $N_{G}=Int(\mathcal{G})+1$ (minimum number of 
channels compatible with $\mathcal{G}$),
Thin solid line:  $N_{G}=Int(\mathcal{G})+2$,  Thin dashed line: 
$N_{G}=Int(\mathcal{G})+3$
The full circles represents present data. The horizontal segments 
represent the temperature dependences of the samples.
     }
\label{minbruit}
   \end{figure}

  From the dimensionless value of the conductance of Rope A, which
varies with $T$ between $\mathcal{G}=G/G_{0}$=11.80 and 
$\mathcal{G}$=11.93 it is possible
to deduce the configuration of $t_i$ which yields the minimum value of $S_I$.
It is obtained assuming that all tubes except 6 are insulator which
means $N_G=12$ conducting channels for the rope. If 5 tubes are perfectly
conducting ($t_i^1 =t_i^2 =1$) and the sixth one is such
that: $t_6^a =1$ then $t_6^b 
=\mathcal{G}-int(\mathcal{G})=\mathcal{G}-11 $. We deduce a shot noise
reduction factor $t_6^b.(1-t_6^b)/\mathcal{G}$ which yields $S_I$ 
between $eI/70$ and $eI/200$
when the resistance varies between 1085~$\Omega$ and 1100~$\Omega$
i.e. of the order of the observed value (see fig.\ref {minbruit}).
It is interesting to consider also the situation in which all the
conducting channels have the same transmission. The maximum noise 
reduction is obtained again when
the number of conducting channels $N_G$ is minimum 
($N_{G}=Int(\mathcal{G}) +1$). Fig.\ref{minbruit}
(thick dashed line) shows that reduction factor peaks for integer 
values of $\mathcal{G}$, which correspond again
to a full transmission ($t_{i}=1$) for all conducting channels.

It is thus in principle nearly possible to
explain the low level of shot noise measured in the ropes by
considering them as a small number ($Int(\mathcal{G})+1$ channels) of 
independent quasi ballistic conductors of non
interacting electrons. For Rope A, this implies for all channels either a
transmission $t_{i}$ equal to 0 or 1 within $\delta t_i$ ($\mid 
\delta t_i \mid <0.1$),
with the very strong additional constraint that $\Sigma_i \mid \delta 
t_i \mid <0.1$
The preliminary results obtained on the individual tube presented in 
fig.5 also indicate
that the level of shot noise is reduced
by more than a factor 3 compared to the maximum shot noise reduction  obtained
assuming that the nanotube consists of 2 independent channels of non
interacting electrons.
As discussed above the presence of electron-electron interactions can
induce a Luttinger liquid state in a 1D system resulting in
non-integer charge excitations
with $e^* = ge$ where g is the LL parameter estimated to be of the
order of 0.3 in carbon nanotubes. Replacing e by e* in (3) could help
fit our experimental reduction of shot noise for all the samples. 
However we want to
stress that the assumption we have made, that only a very small
number of nearly perfectly ballistic conducting tubes contribute to 
transport through
the ropes, implies that the current per tube in our experiments can
exceed $60~\mu A$
per tube within a rope without damaging the sample nor giving
rise to observable non-linearities. This is in apparent contradiction
with previous
experiments both in individual tubes and ropes where non-linearities
start to show up at $5~\mu A$ and the maximum current which can be
sustained by
an individual SWNT is  observed to be of the order of $20~\mu
A$~\cite{yao,avouris}. It is possible that the absence of
non-linearities in our samples is due
to their suspended character: emitted phonons can not leave the tubes 
via the substrate
and are reabsorbed by other electrons as observed in  point
contacts~\cite{kulik}
which can also sustain very high currents.

Assuming instead that more that $Int(\mathcal{G})+1$ channels participate to
transport will yield a shot noise reduction factor which decreases 
very rapidly with $Int(\mathcal{G})+1$.
For example, let us assume that $N_{G}$ conducting channels have the 
same transmission $\mathcal{G}/N_{G}$.
Fig.\ref{minbruit} (thin lines) shows the noise reduction factor for 
$N_{G}=Int(\mathcal{G})+2$ and
$N_{G}=Int(\mathcal{G})+3$.
For all ropes, it falls well below the experimental values. For 
$N_{G}=2*Int(\mathcal{G})$, the effect
is much more drastic with a reduction factor approaching 2.
Such low reductions cannot explain our results. It therefore seems in 
this hypothesis that a much
smaller effective charge is needed to explain the severe reduction we observe.

It may then be necessary to take into account strong correlations between
the current noise in the different tubes constituting the rope or
coupling
between the two channels within each nanotube. The effect of
inter-tubes correlations has been recently discussed theoretically by
Trauzettel {\it et al.}
\cite{trauz} in a more sophisticated experimental setup where coupled
nanotubes can be independently addressed. These authors propose a
scenario with
a perfect locking of the noise in different tubes. Such positive
correlations are expected to increase shot noise. Our experiments
indicate
instead strong anticorrelations.

Note that strong Coulomb repulsion between electrons in vacuum diodes
or triodes have already been shown 60 years ago
\cite{rack} to provide a very efficient mechanism for shot noise reduction.
At this stage both more experimental and theoretical work are
needed to understand our results.

\noindent

We have benifited fom a number of fruitful discussions with Y.
Blanter, M. B\"{u}ttiker, K. Imura, P. Lederer, K.V. Pham and F.
Piechon.
A.K. thanks the Russian foundation for basic research and solid state
nanostructures for financial support, and thanks CNRS for a visitor's
position. P.-E. R. is
grateful to C. Glattli and B. Dou\c{c}ot for their comments.

\end{document}